\documentstyle[12pt]{article}

\def\beq{\begin{equation}}
\def\eeq{\end{equation}}
\def\bea{\begin{eqnarray}}
\def\eea{\end{eqnarray}}
\def\nn{\nonumber}
\def\sss{\scriptscriptstyle}

\def\barp{{\raise.35ex\hbox
{${\sss (}$}}---{\raise.35ex\hbox{${\sss )}$}}}
\def\bdbarp{\hbox{$B_d$\kern-1.4em\raise1.4ex\hbox{\barp}}}
\def\bsbarp{\hbox{$B_s$\kern-1.4em\raise1.4ex\hbox{\barp}}}

\def\roughly#1{\mathrel{\raise.3ex\hbox
{$#1$\kern-.75em\lower1ex\hbox{$\sim$}}}}
\def\lsim{\roughly<}

\def\epjc#1#2#3{{\it Eur.\ Phys.\ J.}\ {\bf C#1}, #3 (19#2)}

\def\npb#1#2#3{{\it Nucl.\ Phys.} {\bf B#1}, #3 (19#2)}
\def\plb#1#2#3{{\it Phys.\ Lett.} {\bf #1B}, #3 (19#2)}
\def\prd#1#2#3{{\it Phys.\ Rev.} {\bf D#1}, #3 (19#2)}

\def\newprdtwo#1#2#3{{\it Phys.\ Rev.} {\bf D#1}: #3 (20#2)}

\def\prl#1#2#3{{\it Phys.\ Rev.\ Lett.} {\bf #1}, #3 (19#2)}
\def\zpc#1#2#3{{\it Zeit.\ Phys.} {\bf C#1}, #3 (19#2)}

\begin{document}
\setlength{\baselineskip}{20pt}

\begin{flushright}
UdeM-GPP-TH-00-70 \\
\end{flushright}

\begin{center}
\bigskip

{\Large \bf T-Odd Triple-Product Correlations}\\
{\Large \bf in Hadronic $b$ Decays}\\
\bigskip

Wafia Bensalem\footnote{wafia@lps.umontreal.ca} and
David London\footnote{london@lps.umontreal.ca}

\medskip

{\it Laboratoire Ren\'e J.-A. L\'evesque, Universit\'e
  de Montr\'eal,}\\
{\it C.P. 6128, succ. centre-ville, Montr\'eal, QC,
  Canada H3C 3J7}\\
\end{center}

\begin{center}
 
\bigskip (\today)

\bigskip 

{\bf Abstract}

\end{center}

\begin{quote}
  We study T-violating triple-product asymmetries in the quark-level
  decay $b\to s u {\bar u}$ within the standard model (SM). We find
  that only two types of triple products are non-negligible. The
  asymmetry in ${\vec p}_u \cdot ( {\vec s}_u \times {\vec s}_{\bar u}
  )$ or ${\vec p}_{\bar u} \cdot ( {\vec s}_u \times {\vec s}_{\bar u}
  )$ can be as large as about 5\%. It can be probed in $B\to V_1 V_2$
  decays, where $V_1$ and $V_2$ are vector mesons. And the asymmetry
  in $\vec s_b \cdot (\vec p_u\times \vec p_s)$ can be in the range
  1\%--3\%. One can search for this signal in the decay $\Lambda_b \to
  \Lambda \pi^+\pi^-$ or in $B^* \to X_s X$, where $X_s$ and $X$ then
  each decay into two mesons. All other triple-product asymmetries are
  expected to be small within the SM.  This gives us new methods of
  searching for new physics.
\end{quote}
\newpage

\section{Introduction}

These are exciting times for $B$ physics. The CDF collaboration has
made a measurement of the CP-violating phase $\sin 2\beta =
0.79^{+0.41}_{-0.44}$ \cite{CDF99}, leading to the nontrivial
constraint $\sin 2\beta > 0$ at 93\% C.L. The asymmetric $e^+ e^-$
$B$-factories BaBar and Belle are now running and will hopefully make
measurements of CP-violating rate asymmetries in the $B$ system before
too long. And in the near future, data from HERA-B and hadron
colliders will add to our knowledge of CP violation in the $B$ system.

The purpose of all this activity is to test the standard model (SM)
explanation of CP violation. In the SM, CP violation, which to date
has been only seen in the kaon system, is due to the presence of a
nonzero complex phase in the Cabibbo-Kobayashi-Maskawa (CKM) quark
mixing matrix $V$. In this scenario, one expects large CP-violating
effects in $B$ decays, and the above experiments are searching for
such signals.

The CP-violating signals which have been the most extensively studied
are rate asymmetries in $B$ decays \cite{CPreview}. Measurements of
such asymmetries will allow one to cleanly probe the interior angles
$\alpha$, $\beta$ and $\gamma$ of the so-called unitarity triangle
\cite{PDG98}, which will in turn provide important tests of the SM.

However, there is another class of CP-violating signals which has
received relatively little attention: triple-product correlations
\cite{Kayser}. In a given decay, it may be possible to measure the
momenta and/or spins of the particles involved. From these one can
construct triple products of the form $\vec v_1 \cdot (\vec v_2 \times
\vec v_3)$, where each $v_i$ is a spin or momentum. Such triple
products are odd under time reversal (T) and hence, by the CPT
theorem, are also potential signals of CP violation. (Note that there
is a technical distinction to be made here: although the action of T
changes the sign of a triple product, if a triple product changes
sign, it is not necessarily due to the T transformation. This is
because, in addition to reversing spins and momenta, the time reversal
symmetry T also exchanges the initial and final states. Thus, in a
particular decay, a nonzero triple product is not necessarily a signal
of T (and CP) violation. For this reason, in what follows we refer to
triple-product asymmetries as {\it T-odd} effects. We also show how to
establish the presence of a true signal of T violation.)

To establish the presence of a nonzero triple-product correlation, one
constructs a T-odd asymmetry of the form
\beq
A_{\sss T} \equiv 
{{\Gamma (\vec v_1 \cdot (\vec v_2 \times \vec v_3)>0) - 
\Gamma (\vec v_1 \cdot (\vec v_2 \times \vec v_3)<0)} \over 
{\Gamma (\vec v_1 \cdot (\vec v_2 \times \vec v_3)>0) + 
\Gamma (\vec v_1 \cdot (\vec v_2 \times \vec v_3)<0)}} ~,
\eeq
where $\Gamma$ is the decay rate for the process in question.
Unfortunately, triple-product correlations suffer from a well-known
complication: their signals can be faked by the presence of strong
phases, even if there is no CP violation. (As noted above, this is
because such correlations are not true T-violating signals.) That is,
one typically finds that
\beq
A_{\sss T} \propto \sin(\phi + \delta),
\eeq
where $\phi$ is a weak, CP-violating phase and $\delta$ is a strong
phase. From this we see that if $\delta \ne 0$, a triple-product
correlation will appear, even in the absence of CP violation (i.e.\ if
$\phi = 0$). 

To remedy this, one can construct the {\it T-violating asymmetry}:
\beq
{\cal A}_{\sss T} \equiv {1\over 2}(A_T-{\bar A}_T) ~,
\label{Tviolasym}
\eeq
where ${\bar A}_{\sss T}$ is the T-odd asymmetry measured in the
CP-conjugate decay process. This is a true T-violating signal in that
it is nonzero only if $\phi \ne 0$ (i.e.\ if CP violation is present).
Furthermore, unlike decay-rate asymmetries in direct CP violation, a
nonzero ${\cal A}_{\sss T}$ does not require the presence of a nonzero
strong phase. Indeed:
\beq
{\cal A}_T\propto \sin {\phi}\cos {\delta} ~,
\eeq
so that the signal is maximized when the strong phase is zero.

As with all CP-violating signals, (at least) two decay amplitudes are
necessary to produce a triple-product correlation. Such correlations
have been studied in semileptonic $B$ decays \cite{Bsemilep}. However,
since there is only a single amplitude in the SM, any such signal can
occur only in the presence of new physics.

To our knowledge, the only study of triple products in the SM has been
made by Valencia \cite{Valencia}, who examined the decay $B \to V_1
V_2$, where $V_1$ and $V_2$ are vector mesons. He looked at triple
products of the form ${\vec k} \cdot ({\vec \epsilon}_1 \times {\vec
  \epsilon}_2)$, where ${\vec \epsilon}_1$ and ${\vec \epsilon}_2$ are
the polarizations of $V_1$ and $V_2$, respectively, and $\vec k$ is
the momentum of one of the vector mesons.  Since the calculation was
done at the meson level, estimates of the various form factors were
needed. The conclusion of this study was that, within the SM, one
could expect a T-violating asymmetry at the level of several percent.

In this paper we re-examine the question of triple products in the SM
using a complementary approach. In particular, we search for
triple-product correlations at the quark level. The motivation is the
following: if a significant triple-product correlation exists at the
hadron level, it must also exist at the quark level. After all, given
that QCD (which is responsible for hadronization) is CP-conserving, it
is difficult to see how one can generate a large T-violating asymmetry
at the hadron level if it is absent at the quark level.

Of course, the converse is not necessarily true: a large T-violating
effect at the quark level might be ``washed out'' during
hadronization, since the spins and momenta of the quarks may not
correlate well with the spins and momenta of the hadrons. (The most
obvious example of this is if spin-0 mesons are involved. In this case
no information about the spins of the constituent quarks can be
obtained.) Thus, there may be considerable hadronic uncertainty in
taking a nonzero quark-level signal and applying it at the hadron
level.

With this in mind, in this paper we examine the inclusive decay $b\to
s u \bar u$ within the SM. If there is a large ${\vec k} \cdot ({\vec
  \epsilon}_1 \times {\vec \epsilon}_2)$ triple product in $B \to V_1
V_2$, there should also be a large triple product at the quark level
of the form ${\vec p} \cdot ({\vec s}\times {\vec s}')$, where ${\vec
  p}$ is the momentum of one of the quarks, and ${\vec s}$ and ${\vec
  s}'$ are the spins of two of the light quarks. And indeed, we find
that the quark-level T-violating asymmetry due to the triple product
${\vec p}_u \cdot ({\vec s}_u \times {\vec s}_{\bar u})$ or ${\vec
  p}_{\bar u} \cdot ({\vec s}_u \times {\vec s}_{\bar u})$ can be as
large as about 5\%. This strongly supports Valencia's conclusion that
the SM predicts a measurable T-violating asymmetry in $B \to V_1 V_2$.

However, we also find another significant T-violating signal in $b\to
s u \bar u$. It is due to the triple-product $\vec s_b \cdot (\vec p_u
\times \vec p_s)$, which involves the $b$-quark spin and the momenta
of the $s$ and $u$ quarks. In the SM, this signal turns out to be in
the range of 1\% to 3\% of the total rate, which may be measurable.
It might be observable in decays such as $\Lambda_b \to \Lambda
\pi^+\pi^-$ (though the usual caveats about large hadronic
uncertainties apply).

Finally, it is also important to note which T-violating signals are
{\it not} present. For example, we find that there are no significant
T-violating asymmetries in the SM which involve the spin of the
$s$-quark. Thus, should such an asymmetry be measured, it would be a
clear sign of new physics.

In Sec.~2, we compute the triple products present in the decay $b\to s
u \bar u$, and estimate their sizes. We discuss possible hadron-level
applications in Sec.~3. We conclude in Sec.~4.

\section{Triple Products in {\lowercase{$b\to s u \bar u$}}}

In the inclusive decay $b\to s u \bar u$, the amplitude has two
dominant contributions: the tree diagram ($T$) due to $W$-boson
exchange and the loop-level strong penguin diagram ($P$). Furthermore,
the penguin amplitude contains two dominant terms, $P_1$ and $P_2$
\cite{Hou}. These various contributions are given by:
\bea
T & = & {4 G_F \over \sqrt{2}} \, V_{ub} V_{us}^* \, 
\Big[ {\bar u} \gamma_\mu \gamma_L b \Big]
\Big[ {\bar s} \gamma^\mu \gamma_L v_u \Big] \, e^{i \delta_t} ~, \nn \\
P_1 & = & -{\alpha_s G_F \over \sqrt{2} \pi} \, F^c_1 \, V_{cb} V^*_{cs} \,
\Big[ {\bar s} t^{\alpha} \gamma_\mu \gamma_L b \Big]
\Big[ {\bar u} t_{\alpha} \gamma^\mu v_u \Big] \, e^{i \delta_1} ~, \nn \\
P_2 & = & -{\alpha_s G_F \over \sqrt{2} \pi} \,
\left[ { -i m_b \over q^2} F_2 \right] \, V_{tb} V^*_{ts} \, 
\Big[{\bar s} t^{\alpha} \sigma_{\mu\nu} q^{\nu} \gamma_R b \Big]
\Big[{\bar u} t_{\alpha} \gamma^\mu v_u \Big] \, e^{i \delta_2}~. 
\eea
In the above, $\gamma_{L(R)} = (1 \mp \gamma_5)/2$, the $t^{\alpha}$
are the Gell-Mann matrices, and the ${\delta}_i$ are the strong
phases. In $P_2$, $q$ is the momentum of the internal gluon. The
factors $F^c_1$ and $F_2$ are functions of $\left(m^2_c/ M^2_W\right)$
and $\left(m^2_t/M^2_W\right)$, respectively, and take the values
$F^c_1 \simeq 5.0$ and $F_2 \simeq 0.2$ for $m_t=160$ GeV \cite{Hou}.
$P_1$ and $P_2$ are often called the {\it chromoelectric dipole
  moment} term and {\it chromomagnetic dipole moment} term,
respectively.

The next step is the calculation of the square of the decay amplitude.
We have:
\beq
{|\cal M|}^2 =  |T|^2 + |P_1|^2 + |P_2|^2 + 2 Re\left(T^\dagger P_1\right)
 + 2 Re\left(T^\dagger P_2\right) + 2 Re\left(P^\dagger_1 P_2\right)~.
\label{ampsquared}
\eeq
We will see below that the dominant term here is $|P_1|^2$. 

We find triple products in all of the interference terms above (i.e.\
the last three terms of ${|\cal M|}^2$). Before giving the specific
forms of these triple products, we make the following general remarks:
\begin{itemize}
  
\item In the calculation, we neglect the masses of the light quarks
  $s$, $u$ and ${\bar u}$, but we keep the spins (i.e.\ polarization
  four-vectors) of these particles (at least to begin with). It turns
  out that there are no triple products involving the polarization of
  the $s$ quark. (In other words, such terms are suppressed by at
  least $m_s/m_b$.) In light of this, in our results below, we
  automatically sum over the $s$-quark spin states.
  
  This is an interesting result: it suggests that if a triple product
  involving the $s$-quark polarization is observed experimentally, it
  is probably due to physics beyond the SM.
  
\item Since the $B$ meson has spin 0, triple products in $B \to V_1
  V_2$ cannot involve the spin of the $b$-quark. If one sums over the
  spin of the $b$-quark, the only term which contains triple products
  is the $T - P_1$ interference term. We will therefore use this term
  to estimate the size of the T-violating asymmetry in $B\to V_1 V_2$
  \cite{Valencia}.
  
\item If the spins of the $u$ and ${\bar u}$ quarks cannot be
  measured, one can then take them to be unpolarized, i.e.\ we sum
  over their polarizations. In this case, only the $T-P_2$ and
  $P_1-P_2$ interferences contain a triple product. This unique signal
  takes the form $\vec s_b \cdot (\vec p_u\times \vec p_s)$.
  
\item In all interference terms, there are triple products which
  involve the three polarizations $\vec s_b$, $\vec s_u$ and $\vec
  s_{\bar u}$. Experimentally, such signals will be extremely
  difficult to measure, and so are of less interest than the others
  described here.

\end{itemize}

\subsection{$T-P_1$ interference}

Keeping explicit the spins of the $b$-, $u$- and ${\bar u}$-quarks,
the T-odd piece of the $T-P_1$ interference term is
\bea
\Bigg[\sum_{ s \ spins} 2Re\left(T^\dagger P_1\right)\Bigg]_{T-odd}&=&
{16{\alpha}_sG^2_FF_1^c
\over 3{\pi}} \,
 Im\bigg[V^{\ast}_{cs}V_{cb}V_{us}V^{\ast}_{ub} 
e^{i\left( {\delta}_1-{\delta}_t\right)}\bigg] \nn\\
&& \hskip-1.5truein \times \, \Bigg\{2(p_b\cdot s_u) \, {\epsilon}_{\mu\nu\rho\xi} \,
p^{\mu}_b p^{\nu}_u p^{\rho}_{\bar u} s^{\xi}_{\bar u} 
- \, 2(p_b\cdot p_u)
\, {\epsilon}_{\mu\nu\rho\xi} \, p^{\mu}_b s^{\nu}_u p^{\rho}_{\bar u} 
s^{\xi}_{\bar u} 
+ \, m^2_b \, {\epsilon}_{\mu\nu\rho\xi} \,
 p^{\mu}_u s^{\nu}_u p^{\rho}_{\bar u} s^{\xi}_{\bar u} \nn\\
&&\hskip-1.2truein 
+m_b\bigg[(s_b\cdot s_u) \, {\epsilon}_{\mu\nu\rho\xi} \,
 p^{\mu}_s p^{\nu}_u p^{\rho}_{\bar u} s^{\xi}_{\bar u} \,
 - \,  (s_b\cdot p_u) \, {\epsilon}_{\mu\nu\rho\xi} \, p^{\mu}_s 
p^{\nu}_{\bar u}  s^{\rho}_{\bar u} s^{\xi}_u \nn\\
&& \hskip-0.8truein 
-(p_s\cdot p_{\bar u}) \, {\epsilon}_{\mu\nu\rho\xi} \, p^{\mu}_u
 s^{\nu}_b s^{\rho}_u s^{\xi}_{\bar u} \, + \, (p_s\cdot s_{\bar u}) \, 
{\epsilon}_{\mu\nu\rho\xi} \, p^{\mu}_u p^{\nu}_{\bar u} s^{\rho}_b
 s^{\xi}_u\bigg]\Bigg\} ~.
\label{TP1int}
\eea
Here, $p_i$ is the 4-momentum of the $i$-quark and $s_i$ is its
polarization four-vector. Triple products\footnote{Note that, due to
  the identity $g_{\alpha \beta} \, \epsilon_{\mu\nu\rho\xi} -
  g_{\alpha \mu} \, \epsilon_{\beta\nu\rho\xi} - g_{\alpha \nu} \,
  \epsilon_{\mu\beta\rho\xi} - g_{\alpha \rho} \,
  \epsilon_{\mu\nu\beta\xi} - g_{\alpha \xi} \,
  \epsilon_{\mu\nu\rho\beta} = 0$, not all terms of the form $v_1
  \cdot v_2 \, \epsilon_{\mu\nu\rho\xi} v^{\mu}_3 v^{\nu}_4 v^{\rho}_5
  v^{\xi}_6$ are necessarily independent.} are found in the terms
$\epsilon_{\mu\nu\rho\xi}v^{\mu}_1 v^{\nu}_2v^{\rho}_3v^{\xi}_4$.

In the above expression, we see that there are two categories of
triple products: those which involve $s_b$, the $b$-quark
polarization, and those which do not. Those terms which include $s_b$
(the last four terms in Eq.~\ref{TP1int}) also include the
polarizations of the $u$- and ${\bar u}$-quarks ($s_u$ and $s_{\bar
u}$). Since all three spins must be measured, these triple products
will be extremely difficult to observe experimentally. Because of
this, it is the first three terms of Eq.~\ref{TP1int} which most
interest us, and we therefore isolate them by averaging over $s_b$.

Of course, as written, the terms $\epsilon_{\mu\nu\rho\xi}v^{\mu}_1
v^{\nu}_2v^{\rho}_3v^{\xi}_4$ involve only four-vectors, and therefore
do not look like triple products. In order to identify the triple
products implicit in these terms, we have to choose a particular frame
of reference. The most natural choice is the rest frame of the
$b$-quark, in which case Eq.~\ref{TP1int} then takes the form
\bea
\Bigg[ {1\over 2} \sum_{b,s \ spins} 2Re\left(T^\dagger P_1\right)\Bigg]_{T-odd}&=&
{16{\alpha}_sG^2_FF_1^c
\over 3{\pi}} \,
 Im\bigg[V^{\ast}_{cs}V_{cb}V_{us}V^{\ast}_{ub} 
e^{i \Delta_{1t} }\bigg] \, m_b^2 \nn\\
&& \hskip-2.0truein \times \, \left\{
s_u^0 \, {\vec p}_u \cdot ( {\vec p}_{\bar u} \times {\vec s}_{\bar u} ) + 
E_u \, {\vec p}_{\bar u} \cdot ( {\vec s}_u \times {\vec s}_{\bar u} ) + 
s_{\bar u}^0 \, {\vec p}_u \cdot ( {\vec p}_{\bar u} \times {\vec s}_u ) + 
E_{\bar u} \, {\vec p}_u \cdot ( {\vec s}_u \times {\vec s}_{\bar u} ) 
\right\} ~,
\label{TP1tps}
\eea
where $\Delta_{1t} \equiv {\delta}_1-{\delta}_t$. We therefore see
that there are, in fact, four distinct triple products in the $T-P_1$
interference term. 

These triple products depend on the polarization four-vectors of the
$u$- and ${\bar u}$-quarks, whose most general form is \cite{KEK}
\beq
s_i^\mu = \left( { {\vec n}_i \cdot {\vec p}_i \over m_i} ~,~
{\vec n}_i + { {\vec n}_i \cdot {\vec p}_i \over m_i (E_i + m_i) } 
{\vec p}_i \right),
\eeq
for $i=u,{\bar u}$. In the above, ${\vec n}_i$ is the polarization
vector of the $i$-quark in its rest frame, and satisfies $|{\vec n}_i|
= 1$. 

The triple products in Eq.~(\ref{TP1tps}) all involve two spins, which
makes their evaluation somewhat problematic. Since one has to choose
directions for ${\vec s}_u$ and ${\vec s}_{\bar u}$, there are an
infinite number of possibilities. We have studied three cases:
\begin{enumerate}
  
\item The polarizations of the $u$- and ${\bar u}$-quarks in their
  respective centre-of-mass frames both point in a particular
  direction, say ${\vec n}_u = {\vec n}_{\bar u} = {\hat z}$.
  
\item The polarizations of the $u$- and ${\bar u}$-quarks are
perpendicular, say ${\vec n}_u = {\hat z}$ and ${\vec n}_{\bar u} =
{\hat x}$.
  
\item The polarization of the $u$-quark is longitudinal, ${\vec n}_u =
  {\hat p}_u$ while ${\vec n}_{\bar u} = {\hat z}$.

\end{enumerate}
In the following we will refer to these three scenarios as Case I,
Case II and Case III, respectively.

In order to compute the size of these triple-product asymmetries, in
addition to integrating over phase space, we also need estimates of
the sizes of the weak and strong phases. In the Wolfenstein
parametrization \cite{Wolfenstein}, we can write the T-odd combination
of CKM and strong phases as
\beq
Im\bigg[V^{\ast}_{cs}V_{cb}V_{us}V^{\ast}_{ub} \, e^{i \Delta_{1t} } \bigg]
= A^2 \lambda^6 \left[ \eta \cos\Delta_{1t} + \rho \sin\Delta_{1t} \right].
\eeq
CP violation in the CKM matrix is parametrized by the parameter
$\eta$. As discussed in the introduction, nonzero strong phases can
fake a T-violating signal. The term $\rho \sin\Delta_{1t}$ in the
above expression is an example of such a fake signal. However, by
forming a true T-violating asymmetry ${\cal A}_{\sss T}$
(Eq.~\ref{Tviolasym}), one can eliminate this fake signal. In this
case ${\cal A}_{\sss T} \propto \eta \cos\Delta_{1t}$.

At present, $\eta$ is constrained to lie in the range $0.23 \le \eta
\le 0.50$ \cite{AliLon}. Turning to the strong phase, the tree-level
phase $\delta_t$ is usually assumed to be small: the logic is that,
roughly speaking, the quarks will hadronize before having time to
exchange gluons. On the other hand, for the $b\to s u \bar u$ penguin
amplitude, it is often assumed that strong phases come from the
absorptive part of the penguin contribution \cite{bssgh}. Since $P_1$
involves an internal $c$-quark, it is possible that $\delta_1 \ne 0$,
which of course implies that $\Delta_{1t} \ne 0$. Even so, for
simplicity, in our calculation we assume that $\Delta_{1t}$ is small
enough that $\cos\Delta_{1t} \simeq 1$ is a good approximation.
However, the reader should be aware that the asymmetries may be
reduced should this strong phase be large. (Note that the T-violating
signal is maximal when $\cos\Delta_{1t} = 1$. For comparison, direct
CP-violating rate asymmetries require the strong phase to be nonzero.)

We have performed the phase-space integration using the computer
program RAMBO. For each of the three cases above we have calculated
the four T-violating asymmetries [see Eq.~(\ref{Tviolasym})] ${\cal
  A}_{\sss T}^1$, ${\cal A}_{\sss T}^2$, ${\cal A}_{\sss T}^3$, and
${\cal A}_{\sss T}^4$, which correspond respectively to the four
triple products of Eq.~(\ref{TP1tps}): $s_u^0 \, {\vec p}_u \cdot (
{\vec p}_{\bar u} \times {\vec s}_{\bar u} )$, $E_u \, {\vec p}_{\bar
  u} \cdot ( {\vec s}_u \times {\vec s}_{\bar u} )$, $s_{\bar u}^0 \,
{\vec p}_u \cdot ( {\vec p}_{\bar u} \times {\vec s}_u )$, and $E_{\bar
  u} \, {\vec p}_u \cdot ( {\vec s}_u \times {\vec s}_{\bar u} )$. We
have taken $\eta = 0.4$. Our results are as follows.
\begin{enumerate}
  
\item Case I: The asymmetries ${\cal A}_{\sss T}^1$ and ${\cal
A}_{\sss T}^3$ are both negligible. However, we find that ${\cal
A}_{\sss T}^2 = {\cal A}_{\sss T}^4 \simeq 4.6\%$.

\item Case II: ${\cal A}_{\sss T}^1$ and ${\cal A}_{\sss T}^3$ are
negligible, while ${\cal A}_{\sss T}^2 = {\cal A}_{\sss T}^4 \simeq
3.9\%$.
  
\item Case III: Here the triple products ${\vec p}_u \cdot ( {\vec
p}_{\bar u} \times {\vec s}_u )$ and ${\vec p}_u \cdot ( {\vec s}_u
\times {\vec s}_{\bar u} )$ vanish identically, so that ${\cal
A}_{\sss T}^3 = {\cal A}_{\sss T}^4 = 0$. We find that the other two
asymmetries are tiny: ${\cal A}_{\sss T}^2 = - {\cal A}_{\sss T}^1
\simeq 0.09\%$.

\end{enumerate}
{}From the above, we conclude that the asymmetries ${\cal A}_{\sss T}^1$
and ${\cal A}_{\sss T}^3$ are both very small in the SM, and that
${\cal A}_{\sss T}^2$ and ${\cal A}_{\sss T}^4$ can be as large as
about 5\%.

Note that the triple product in $B\to V_1 V_2$ discussed by Valencia
\cite{Valencia} is of the form ${\vec k} \cdot ({\vec \epsilon}_1
\times {\vec \epsilon}_2)$. In Eq.~(\ref{TP1tps}), it is the terms
${\vec p}_{\bar u} \cdot ( {\vec s}_u \times {\vec s}_{\bar u} )$ and
${\vec p}_u \cdot ( {\vec s}_u \times {\vec s}_{\bar u} )$ which could
potentially give such a triple-product signal. We have found that the
asymmetries ${\cal A}_{\sss T}^2$ and ${\cal A}_{\sss T}^4$, which
correspond to these triple products, can be reasonably big ($\lsim
5\%$). This is consistent with the results found by Valencia at the
meson level, and suggests that the SM does indeed predict a measurable
T-violating asymmetry in $B\to V_1 V_2$ decays.  (Of course, due to
our difficulties in understanding hadronization, it is not possible to
use the quark-level result to predict the size of the asymmetry for a
specific meson-level decay.)

Finally, for comparison, consider the decay-rate asymmetry, calculated
by Hou for the same process \cite{Hou2}:
\beq
a_{CP}(b\to s u \bar u)\simeq 1.4\%
\eeq
We therefore see that one expects T-violating triple-product
asymmetries in $b\to s u \bar u$ to be considerably larger than the
decay rate asymmetry.

\subsection{$P_1-P_2$ interference}

The T-odd piece of the $P_1-P_2$ interference term is
\bea
\Bigg[\sum_{s \ spins} 2Re\left(P^\dagger_1P_2\right)\Bigg]_{T-odd}&=&
{4{\alpha}^2_sG^2_FF^c_1
F_2m_b\over 3{\pi}^2q^2}\, Im\bigg[V^{\ast}_{ts}V_{tb}V_{cs}V^{\ast}_{cb}
\, e^{i \left({\delta}_2- {\delta}_1\right)}\bigg]\nn\\
&& \hskip-1.5truein 
\times \, \Bigg\{\Big[p_b \cdot \left(p_u-p_{\bar u}\right) (1-s_u \cdot 
s_{\bar u})
 \, - \, (s_{\bar u} \cdot p_s)(s_u\cdot p_{\bar u})
+ \, (s_u\cdot p_s)(s_{\bar u}\cdot p_u)\Big] \, 
 {\epsilon}_{\mu\nu\rho\xi} \, p^{\mu}_b s^{\nu}_b p^{\rho}_u p^{\xi}_s \nn\\
&& \hskip-1.2truein 
+ \, \Big[(s_u\cdot p_{\bar u})(p_u\cdot p_s)-{q^2\over  2}
 (s_u\cdot p_s)\Big] \, {\epsilon}_{\mu\nu\rho\xi} \, 
 p^{\mu}_b s^{\nu}_b p^{\rho}_s s^{\xi}_{\bar u} \nn\\
&& \hskip-1.2truein 
+ \, \Big[(s_{\bar u}\cdot p_u)(p_{\bar u}\cdot p_s)-{q^2\over  2}
 (s_{\bar u}\cdot p_s)\Big] \, {\epsilon}_{\mu\nu\rho\xi} \, 
 p^{\mu}_b s^{\nu}_b p^{\rho}_s s^{\xi}_u \,   
\Bigg\} ~.
\label{P1P2int}
\eea
Here, if we average over the $b$-quark spin states, there is no
T-violating signal at all.

We note that most of the terms in Eq.~\ref{P1P2int} correspond to
triple products in which three spins must be measured. As we have
already discussed, such signals are very difficult to observe
experimentally, and so do not interest us. There is one term, however,
which does not involve three spins, and it can be isolated by summing
over the $u$- and ${\bar u}$-quark spin states:
\bea
\Bigg[\sum_{u, \bar u, s \ spins} 2Re\left(P^\dagger_1P_2\right)\Bigg]_{T-odd}
 &=&{16{\alpha}^2_sG^2_FF^c_1
F_2m_b\over 3{\pi}^2q^2}\, Im\bigg[V^{\ast}_{ts}V_{tb}V_{cs}V^{\ast}_{cb}
\, e^{i \left({\delta}_2- {\delta}_1\right)}\bigg]\nn\\
&&\times \, p_b \cdot \left(p_u-p_{\bar u}\right) \,
 {\epsilon}_{\mu\nu\rho\xi} p^{\mu}_b s^{\nu}_b p^{\rho}_u p^{\xi}_s ~.
\eea
In the rest frame of the $b$-quark, the triple product takes the form
$m_b^2 (E_u - E_{\bar u}) \, \vec s_b \cdot (\vec p_u\times \vec
p_s)$. Integrating over phase space with RAMBO, we find that the
$P_1-P_2$ T-violating asymmetry is $O(10^{-5})$, which is negligible.

\subsection{$T-P_2$ interference}

Like $P_1-P_2$ interference, the $T-P_2$ interference term contains
two types of triple produts: (i) those involving a single quark
polarization, $s_b$, and (ii) those involving the three polarization
four-vectors $s_b$, $s_u$ and $s_{\bar u}$. As usual, we are not
interested in triple products involving three spins, and so we can
therefore sum over $s_u$ and $s_{\bar u}$. The T-odd piece of the
$T-P_2$ interference term is then given by
\bea
\Bigg[\sum_{u, \bar u, s \ spins} 2Re\left(T^\dagger P_2\right)\Bigg]_{T-odd}
&=&{128{\alpha}_sG^2_FF_2m_b
\over 3{\pi}q^2} \, Im\bigg[V^{\ast}_{ts}V_{tb}V_{us}V^{\ast}_{ub} \,
e^{i \Delta_{2t} }\bigg]\nn\\
&&\times \, p_s \cdot p_u \, {\epsilon}_{\mu\nu\rho\xi} \,
p^{\mu}_b s^{\nu}_b p^{\rho}_u p^{\xi}_s ~,
\label{TP2asym}
\eea
where $\Delta_{2t} \equiv \delta_2 - \delta_t$.

As was the case for the $T-P_1$ interference term, the T-violating
asymmetry ${\cal A}_{\sss T}$ is proportional to $\eta
\cos\Delta_{2t}$. And, as before, we expect the $\delta_t$ piece of
the strong phase $\Delta_{2t}$ to be small. However, there is a
difference here compared to the $T-P_1$ case: previously, the penguin
amplitude $P_1$ involved an internal $c$-quark, and so it was possible
that the strong phase $\delta_{1t}$, which is related to the
absorptive part of the amplitude, could be sizeable. Here, the triple
product involves only the $t$-quark penguin contribution $P_2$, which
is purely dispersive, and so leads to $\delta_2=0$. Thus, it is an
excellent approximation to set $\Delta_{2t} \simeq 0$.

In the rest frame of the $b$-quark, the triple-product of
Eq.~\ref{TP2asym} is $m_b \, p_s \cdot p_u \, {\vec s}_b \cdot ( {\vec
p}_u \times {\vec p}_s )$. Integrating over phase space using RAMBO,
and using the allowed range for $\eta$, we find that this T-violating
triple-product asymmetry can be of the order of several percent:
\beq
1.3\% \lsim {\cal A}_T(b\to s u \bar u) \lsim 3.2\% ~.
\eeq
This could conceivably be measured at a future experiment.

Furthermore, if it is found that this asymmetry is considerably larger
than the above values, it is probably a signal of new physics. For
example, in some models of new physics, the chromomagnetic dipole
moment $F_2$ can be enhanced up to ten times its SM value
\cite{F2enhance}. This will clearly have an enormous affect on the
above asymmetry.

\section{Applications}

In the previous section, in our study of the quark-level decay $b\to s
u {\bar u}$ within the SM, we found two classes of triple products
whose T-violating asymmetry is large. They are: (i) $E_u \, {\vec
  p}_{\bar u} \cdot ( {\vec s}_u \times {\vec s}_{\bar u} )$ and
$E_{\bar u} \, {\vec p}_u \cdot ( {\vec s}_u \times {\vec s}_{\bar u}
)$, and (ii) $m_b \, p_s \cdot p_u \, {\vec s}_b \cdot ( {\vec p}_u
\times {\vec p}_s )$. The next obvious question is then: how can one
test these results?

The ideal way would be to make triple-product measurements {\it
  inclusively}. If this were possible, then it would be
straightforward to compare the experimental values with the
theoretical predictions. However, this may not be experimentally
feasible, in which case we must turn to exclusive $B$ decays.

The first class of triple-product asymmetries can be studied in $B\to
V_1 V_2$ decays which are dominated by the quark-level process $b\to s
u {\bar u}$. Examples of such decays include ${\overline{B_d^0}} \to
\rho K^*$, ${\overline{B_s^0}} \to K^{*+} K^{*-}$, $B_c^- \to D^* K^{*-}$,
etc. These have been examined by Valencia, and we refer the reader to
Ref.~\cite{Valencia} for details.

Turning to the second class of triple products, it is clear that we
cannot use decays of $B$ mesons to obtain these asymmetries: since the
$B$-meson spin is zero, there is no way to measure the spin of the
$b$-quark (which is the only spin contributing to the triple product).
However, one possibility would be to use the $\Lambda_b$ baryon, whose
spin is largely that of the $b$ quark. For example, we can consider
the process $\Lambda_b \to \Lambda \pi^+\pi^-$. The triple product
${\vec s}_b \cdot ( {\vec p}_u \times {\vec p}_s )$ can be roughly
equated to $\vec s_{\Lambda_b} \cdot(\vec p_{\pi^+} \times \vec
p_{\Lambda})$.

Another possibility is to consider a $B^*$ meson decaying into any two
mesons $X_s X$, where $X_s$ and $X$ then decay respectively into
mesons $\Phi_1\Phi_2$ and $\Phi_3\Phi_4$ (since with only one spin, we
need three independent momenta). The triple product $\vec s_b \cdot
(\vec p_u\times \vec p_s )$ may then be roughly related to $\vec
s_{B^{\ast}} \cdot(\vec p_{\Phi_1}\times \vec p_{\Phi_3})$.

In the above, we have been deliberately vague about the relationship
between the triple products at the quark and hadron levels. We do not
understand hadronization of quarks into mesons all that well, and when
one adds the complication of relating the quark spins to the hadron
spins, things are even more uncertain. Given that there is a large
quark-level triple-product asymmetry, there may be one at the hadron
level. However, if experiment does not find such an asymmetry, this
may not be a sign of new physics -- it may simply mean that the
asymmetry has been washed out during hadronization. Regardless of the
results, studies of this kind are likely to help us understand how
quarks hadronize into mesons and baryons.

Finally, we note that certain quark-level triple products are
predicted to be small in the SM. For example, triple products
involving the spin of the $s$-quark are suppressed by powers of its
mass. Hence, if a T-violating asymmetry due to a triple product
involving the $s$-quark spin were found to be sizeable, this would
probably indicate the presence of new physics. The decay $\Lambda_b
\to \Lambda \pi^+\pi^-$, which was mentioned above, can be used to
test this. The spin of the $\Lambda$ is due mostly to the $s$-quark
spin. So any T-violating asymmetry involving the spin of the
$\Lambda$, such as $\vec s_{\Lambda_b} \cdot(\vec s_{\Lambda} \times
\vec p_{\Lambda})$, should be tiny in the SM.

As another example, recall that we found that $T-P_1$ interference
produced the triple products $s_u^0 \, {\vec p}_u \cdot ( {\vec
  p}_{\bar u} \times {\vec s}_{\bar u} )$ and $s_{\bar u}^0 \, {\vec
  p}_u \cdot ( {\vec p}_{\bar u} \times {\vec s}_u )$. However, the
corresponding T-violating asymmetries ${\cal A}_{\sss T}^1$ and ${\cal
  A}_{\sss T}^3$ turned out to be suppressed dynamically. Consider the
decay of a $B$-meson to two vector mesons, $B \to V_1 V_2$, where the
$V_2$ then subsequently decays to two mesons $\Phi_1\Phi_2$. Roughly
speaking, one can relate $s_u^0 \, {\vec p}_u \cdot ( {\vec p}_{\bar
  u} \times {\vec s}_{\bar u} )$ to ${\epsilon}_{V_1}^0 \, {\vec
  p}_{V_1} \cdot ( {\vec \epsilon}_{V_2} \times {\vec p}_{\Phi_1} )$.
Thus, the measurement of a nonzero value for this latter
triple-product asymmetry would be a signal for new physics.

\section{Conclusions}

We have calculated the quark-level triple-product correlations in the
decay $b \to s u {\bar u}$ within the standard model. Although several
such triple products are present, we find that only two types lead to
sizeable T-violating asymmetries. 

The first type includes ${\vec p}_u \cdot ( {\vec s}_u \times {\vec
s}_{\bar u} )$ and ${\vec p}_{\bar u} \cdot ( {\vec s}_u \times {\vec
s}_{\bar u} )$.  We find that the corresponding T-violating
asymmetries can be as large as about 5\%. This triple product can be
probed in $B\to V_1 V_2$ decays, where $V_1$ and $V_2$ are vector
mesons \cite{Valencia}.

The second type is $\vec s_b \cdot (\vec p_u\times \vec p_s)$, where
${\vec s}_b$ is the polarization of the $b$-quark, and $\vec p_u$ and
$\vec p_s$ are the momenta of the $u$- and $s$-quark, respectively. We
calculate that the T-violating asymmetry for this triple product is in
the range 1\%--3\%, which may be measurable. There are several ways to
try to search for this triple-product asymmetry. For example, one
could study the decay $\Lambda_b \to \Lambda \pi^+\pi^-$, looking for
a nonzero triple product $\vec s_{\Lambda_b} \cdot(\vec p_{\pi^+}
\times \vec p_{\Lambda})$. Another possibility is to examine the decay
$B^* \to X_sX$, where $X_s$ and $X$ then decay respectively into
mesons $\Phi_1\Phi_2$ and $\Phi_3\Phi_4$, and to search for the triple
product $\vec s_{B^{\ast}} \cdot(\vec p_{\Phi_1}\times \vec
p_{\Phi_3})$.

The fact that we find only two large triple-product correlations has
interesting consequences. If a triple product is tiny at the quark
level, it is probably tiny at the hadron level as well. After all, the
hadronization of quarks into hadrons is a strong-interaction process,
and QCD is CP-conserving. It is therefore difficult to see how one can
generate a large triple-product correlation at the hadron level, given
that it is small at the quark level. (Of course, the converse is not
necessarily true: it is quite possible that a quark-level CP-violating
effect might be ``washed out'' during hadronization.)

{}From the point of view of looking for physics beyond the SM, it is
therefore important to identify those triple-product asymmetries which
are expected to be small in the SM. If such asymmetries are found to
be large, this is probably a signal of new physics. For example, we
find that triple products involving the spin of the $s$-quark are
suppressed by powers of its mass. Thus, if, for instance, a sizeable
T-violating asymmetry of the form $\vec s_{\Lambda_b} \cdot(\vec
s_{\Lambda} \times \vec p_{\Lambda})$ were found in the decay
$\Lambda_b \to \Lambda \pi^+\pi^-$, this would be compelling evidence
for the presence of new physics, since the spin of the $\Lambda$ is
due largely to the $s$-quark spin.

Note that we have performed the calculation at the quark level, and
the passage from quarks to hadrons is not well understood. Thus, in
addition to being an interesting signal of T and CP violation, the
study of T-violating triple-product asymmetries may help us understand
aspects of hadronization.

\section{\bf Acknowledgments}

W.B. would like to thank P. Depommier for helpful discussions, and G.
Azuelos for help with RAMBO. D.L. thanks B. Kayser and A. Soni for
useful conversations. This work was financially supported by NSERC of
Canada.


\begin{thebibliography}{99}
  
\bibitem{CDF99} CDF Collaboration, T. Affolder et al.,
  \newprdtwo{61}{00}{072005}.

\bibitem{CPreview} For a review, see, for example, {\it The BaBar
    Physics Book}, eds.\ P.F. Harrison and H.R. Quinn, SLAC Report
  504, October 1998.

\bibitem{PDG98} C. Caso et al.\ (Particle Data Group),
  \epjc{3}{98}{1}.
 
\bibitem{Kayser} A general discussion of triple products in $B$ decays
  can be found in B. Kayser, {\it Nucl.\ Phys.\ B (Proc.\ Suppl.)}
  {\bf 13}, 487 (1990).

\bibitem{Bsemilep} E. Golowich and G. Valencia, \prd{40}{89}{112};
  J.G. K\"orner, K. Schilcher and Y.L. Wu, \plb{242}{90}{119},
  \zpc{48}{90}{663}; G.-H. Wu, K. Kiers and J.N. Ng,
  \plb{402}{97}{159}, \prd{56}{97}{5413}.
  
\bibitem{Valencia} G. Valencia, \prd{39}{89}{3339}.

\bibitem{Hou} W.-S. Hou, \npb{308}{88}{561}.
  
\bibitem{KEK} See, for example, KEK Experiment E246, Technical Note
  No.~28.
  
\bibitem{Wolfenstein} L. Wolfenstein, \prl{51}{83}{1945}.

\bibitem{AliLon} A. Ali and D. London, {\it hep-ph}/0002167.

\bibitem{bssgh} M. Bander, D. Silverman and A. Soni, \prl{43}{79}{242};
  J.-M. Gerard and W.-S. Hou, \prd{43}{91}{2909}

\bibitem{Hou2} W.-S. Hou, {\it hep-ph}/9902382.
  
\bibitem{F2enhance} For example, see A. Kagan, hep-ph/9806266; W.S.
  Hou, hep-ph/9902382, and references therein.

\end{thebibliography}
\end{document}